# Can Artificial Intelligence Make Art?

## Folk Intuitions as to whether AI-driven Robots Can Be Viewed as Artists and Produce Art

Elzė Sigutė Mikalonytė *
Institute of Philosophy, Vilnius University, elze.mikalonyte@fsf.vu.lt

Markus Kneer
Department of Philosophy, University of Zurich, markus.kneer@uzh.ch

In two experiments (total N=693) we explored whether people are willing to consider paintings made by AI-driven robots as *art*, and robots as *artists*. Across the two experiments, we manipulated three factors: (i) agent type (AI-driven robot v. human agent), (ii) behavior type (intentional creation of a painting v. accidental creation), and (iii) object type (abstract v. representational painting). We found that people judge robot paintings and human paintings as art to roughly the same extent. However, people are much less willing to consider robots as artists than humans, which is partially explained by the fact that they are less disposed to attribute artistic intentions to robots.

**CCS CONCEPTS** • Human-centered computing ~Human computer interaction (HCI) ~Empirical studies in HCI • Applied Computing ~Arts and humanities ~Fine arts • Computing methodologies ~Artificial Intelligence ~ Philosophical/theoretical foundations of artificial intelligence ~Cognitive science

**ACM Reference Format:**
Mikalonytė, E. S., and Kneer, M. 2022. Can Artificial Intelligence Make Art? ACM Transactions on Human-Robot Interaction.

---

* The authors contributed equally to this publication. Kneer is also affiliated with the Digital Society Initiative (University of Zurich).

# 1 Introduction

Nowadays, encounters with works of art created by robots are no longer unusual: We come across paintings, poems, songs and even film scripts produced by machines. These works are presented to us in the usual ways: Robots have exhibitions [2] and their works are bought for hundreds of thousands of dollars [15]. And yet, when it comes to the question whether robot-created objects are genuine works of art, opinions diverge. Art has traditionally been considered to be one of those domains exclusive to humans, as creativity – sometimes called "the final frontier" of AI research [17] – is highly valued by society, and is not that easily attributed to non-human entities, especially those which do not have mental states. There is a reasonable doubt whether robots can really create art.

If, in our effort to answer this question, we turn to (an analogue of) the classic Turing test [84], the answer appears straightforward: We already know that artificial intelligence can produce art-like works that are not only indistinguishable from those produced by human agents, but that these works are also perceived as having no less of an aesthetic value [28, 9]. However, many believe that robots cannot create art. Common reasons include that machines do not have human-like intelligence, autonomy, mental states, emotions or – partially as a consequence of the latter – the agency necessary to participate in social relations [35, 36].

If, however, one were to deem (certain types of) robot-made creations art, a second interesting question arises: Are we willing to consider robots as artists? And if so, under what conditions? One might distinguish, following d'Inverno and McCormack [27], between "Heroic AI" and "Collaborative AI". The former refers to independent creative autonomous agents, the latter to AI which is part of a group agent that includes humans. Naturally, there are also non-autonomous machines designed to be used as mere tools in the process of art creation. In the literature on robot ethics, there are proposals to attribute collective agency to human-robot collaborations [67] and this proposal could be extended to aesthetic agency as well. Others, however, attribute the authorship of the non-human created artworks solely to the human authors of the creative machines [80]. Just as in moral contexts, where we feel uncertain of who is responsible for robot behavior [79, 13, 58], in aesthetics we might feel unsure who is responsible for artistic robot creation.

The authorship problem is also reflected in the legal discussion on the ownership of the copyrights of the artworks produced by robots. Although works of art that are autonomously created by AI are currently not copyrighted, there are proposals to assign authorship to robots by redefining the expression "authorship" to include non-human agents [22, 1, 75], while others propose to transfer the copyrights of the robot-created works to their human designers [10, 39].

## 1.1 Folk intuitions about art made by AI

To what extent are people willing to call robots artists and their creations art? Although there is agreement among researchers that machines do not have consciousness [24, 37], people have a strong tendency to anthropomorphise robots and tend to ascribe a wide variety of mental states to them (for a review, see [69], as well as the references in section 1.2). While a lot of recent research has been conducted on intuitions regarding the perceived moral agency of robots, the same cannot be said about intuitions regarding the ability of robots to create art, and this area remains underexplored. Although some claim that "AI systems are not broadly accepted as authors by artistic or general public communities" [59], there is currently almost no empirical research on folk intuitions (one exception is [30]).

Even if we discover that folk intuitions suggest that robots do create art and have all the mental states needed to be real creators, it won't mean that those intuitions should be decisive in either philosophical or legal discussions. On the contrary, many authors see the tendency to anthropomorphise AI as problematic [74] and caution care about the ascription of rich psychological states to robots [77]. In the same vein, some authors claim that it would be "misleading and risky" not to raise awareness of this tendency to project human traits onto AI [76] Naturally, to better understand the folk propensity to ascribe agentic traits to robots, more empirical research is needed and we should look beyond moral contexts only.

How should empirical research on intuitions about robots' ability to create art be conducted? If we examine the positions on the possibility of the robot-created art from the philosophical point of view, it is helpful to consider Mark Coeckelbergh's [14] proposal to break the question "can machines create art?" down into three smaller ones: What do we mean by creation, by art, and by machines? In trying to determine whether robots are perceived as being able to create art, we need to single out three aspects: the *agent* (e.g. an autonomous robot, as compared to human), the *process* (the action by which a work is brought to life), and the *product* (the created object).

All three factors are important from the philosophical point of view and they are discussed at length in Coeckelbergh's paper [14]. The first factor, or the agent, raises questions about how we should think about various types of art creators – humans, robots, or human-robot teams. This factor also invites reflection on whether the purpose of robot artists is to imitate humans. The second factor, or the process, raises the question of whether machines can engage in a genuine creative process or whether their movements are "just movements". The main question regarding *creativity* is whether it presupposes the existence of the creator's consciousness and mental states. Finally, the third factor, or the product, is important because it relates to the more general problem of defining art: Can art be defined, for instance, in terms of its aesthetic properties or institutional recognition? Is robot-created "art" compatible with traditional philosophical definitions of art? We will discuss these questions in further detail in the following sections.

## 1.2 Mental state ascription

Many scholars define art in ways that depend on the creator's mental states. A classic account is Jerrold Levinson's intentional-historical definition of art, which states that:

> "X is an artwork at [time] t = df X is an object of which it is true at t that some person or persons, having the appropriate proprietary right over X, non-passingly intends (or intended) X for regard-as-a-work-of-art, i.e., regard in any way (or ways) in which objects in the extension of 'artwork' prior to t are or were correctly (or standardly) regarded." [47]

Beardsley's aesthetic definition of art also stresses the creator's intention, although it does not treat it as a necessary condition for an object to be considered a work of art. It states that a work of art is

> "[...] either an arrangement of conditions intended to be capable of affording an experience with marked aesthetic character or (incidentally) an arrangement belonging to a class or type of arrangements that is typically intended to have this capacity." [7]

There is also a romantic conception of the creation of art, according to which the latter is an expression of the creator's inner world and the creator's emotions [16]. All these conceptions of art are hard to reconcile with the position that robots can make art, given that they do not have intentions and there isn't much of an inner world to express.

As we know from research in the psychology of art, inferences about creators' mental states play an important role in our reasoning about artworks. Paul Bloom (who draws inspiration from Levinson's intentional-historical definition of art) claims that people reason about artworks (and all artifacts) in terms of inferred authorial intent. They categorize an object as a member of the artifact kind if it is thought to be created with the intention for it to belong to that kind:

> "We infer that a novel entity has been successfully created with the intention to be a member of artifact kind X – and thus is a member of artifact kind X – if its appearance and potential use are best explained as resulting from the intention to create a member of artifact kind X." [8]

Inferences about the mental states of creators are as important for categorizing works of art as for other artifacts. According to Bloom, authorial intent is relevant to determining the kind of an artifact even if it is not directly connected to the object's appearance or function. Thus, intuitions as to whether an object was created with an intention for it to belong to the category of artworks can play an important part in judgments as to whether an object should be categorized as art. Consequently, people's willingness to consider robot-created objects art is contingent on their disposition to attribute intentions to robots in the first place.

Recent empirical work confirms Bloom's proposal: When judging whether an object falls under the category of "artwork", the intent of the creator is seen as more important than even the appearance of the object in question [66]. Jucker et al. [40] found that perceived artists' intentions affect what people categorize as art, as well as evaluations of what is good art (see also [65, 70], and see [5] for Non-Western cultures; also see [12] for the influence of perceived anthropomorphism on aesthetic responses to computer-generated works of art). Children, too, manifest the tendency to categorize objects as art and interpret them according to the perceived artists' intentions [32, 33, 71].

Are people willing to ascribe intentionality to robots? The folk concept of intentionality consists of five elements: (i) a desire for an outcome, (ii) a belief about the action leading to that outcome, (iii) an intention to perform the action, (iv) awareness of fulfilling the intention while performing the action, and (v) the skill to perform the action [54]. Empirical research shows that all these elements are ascribed to robots, often (though not across the board) to similar degrees as to humans (see [69] for a review). People are, for instance, willing to attribute inculpating mental states in trolley dilemmas to AI-driven robots [85], foreknowledge of harm [81], or recklessness to robots in contexts of risk [45]. They are disposed to attribute intentions to robots [83], such as the intention to deceive and the capacity to lie ([44], see also [86] and [26]). When given a choice between mentalistic and mechanistic vocabulary to explain robot action, people are quite willing to use the former [57] and they invoke similar concepts to describe robot action as for human action [23]. Furthermore, fMRI studies have shown that perceived robot action gives rise to cortical activity related to mental state inferences [46, 68]. Since people are willing to attribute mental states to robots, it does not surprise that, in contexts of harm, they ascribe moral responsibility to them [29, 87], are willing to blame them (see e.g. [55, 56, 45]) and want to punish them [49] – even if these machines only have "a body to kick, but still no soul to damn" [4]. Given the core importance of perceived mental states for the *moral* evaluation of robot actions and robot agents, we can expect them to play a similar role in the *aesthetic* evaluation of robot creations and creative agency.

It should be noted that there are conceptions of art which do not require artworks to have been created by agents with mental capacities. For example, George Dickie's institutional definition of art describes a work of art as something which is:

> "(1) an artifact (2) a set of the aspects of which has had conferred upon it the status of candidate for appreciation by some person or persons acting on behalf of a certain social institution (the art-world)" [25]

According to this view, institutional recognition would be enough for us to acknowledge a robot's creation as a work of art. The classical mimetic theory of art [3, 34] is another conception which does not necessarily invoke mental states on the part of the artwork's creator. If we consider art to be an imitation of the outer world, as opposed to an expression of the creator's inner world, robots might well be deemed creative. The empirical studies presented below will thus not only shed light on the question of whether the folk are willing to consider robot creations as art. They will also help elucidate the question of what the folk concept of art is more generally – and in particular, whether the creation of art is tied to an intention to make art.

## 1.3 Can AI make art?

In our empirical studies, we wanted to explore the question whether people are disposed to see paintings made by robots as art and whether they consider robots as artists. Some work employing the Computers Are Social Actors (CASA) paradigm [38, 64] predicts we do, since heuristics at work in human-human interactions are unreflectively extended to human-robot interactions. Dovetailing with Coeckelbergh's [14] three core factors pertaining to this debate, we manipulated (i) *agent* type (human v. AI-driven, autonomous robot), (ii) *process* type (intentional v. accidental creation), and (iii) *product* type (abstract v. representational painting). In light of the foregoing discussion concerning the importance of perceived mental capacity, we tested the extent to which people are willing to ascribe a belief, a desire and an intention to the robot agent to make a painting. Our core questions were:

> Q1: Are people as willing to judge a robot creation art as a human creation?
> Q2: Are people as willing to judge AI-driven robot creators artists as human creators?
> Q3: Are people willing to ascribe artistic intentions to AI-driven robot agents, and to what extent can the latter elucidate the findings regarding Q1 and Q2?

Experiment 1 reports the data for a scenario in which the agent produced an *abstract* painting. Experiment 2 reports the data for a scenario in which the agent produced a *representational* painting, i.e. a kind of work that might increase inferences concerning the agent's intentionality.

## 2 Experiment 1 - Abstract Art

### 2.1 Participants

We recruited 392 participants online via Amazon Mechanical Turk. The IP address location was restricted to the USA. In line with the preregistered criteria,[1] we excluded participants who failed an attention check or took less than eight seconds to answer the questions (not including reading the scenario), leaving a sample of 254 participants (female: 53%; age M=44 years, SD=14 years, range: 23–79 years).

### 2.2 Methods and Materials

Participants were shown a vignette (see Appendix for detail) in which either a human, or else an autonomous, AI-driven robot creates an abstract painting. Besides agent type, we manipulated how the painting came about: In one condition, the agent decides to make a painting (intentional), in the other condition they clean up the studio, and accidentally knock over paint that spills onto a canvas (accidental). In total there were thus four conditions (2 agent types: human v. robot x 2 behavior types: intentional v. accidental), to which participants were assigned randomly. The *robot / intentional* condition, for instance, read:

> "Imagine a robot equipped with artificial intelligence that can make decisions autonomously. The robot is in an art studio and decides to create a piece of art. It takes an empty canvas and starts splashing paint onto it. In the end, the object looks like an abstract painting."

Having read the scenario, participants had to rate on a Likert scale anchored at 1 with "completely disagree" and 7 with "completely agree" to what extent they agreed with the following claims (labels in brackets omitted):

(1) "The painting is art" (Art)
(2) "The painting was made by an artist" (Artist)
(3) "The agent wanted to make a painting" (Desire)
(4) "The agent believed they were making a painting" (Belief)
(5) "The agent intentionally made a painting" (Intention)

Once finished with the task, participants had to complete a demographic questionnaire.

### 2.3 Results

#### 2.3.1 Main Results.

A series of mixed-design ANOVAs determined that, aggregating across the two behavior type conditions, participants were more willing to confer the status of art to the human's painting than to the robot's ($F(1,253)=7.598$, $p=.006$, $\eta_p^2=.029$). However, the effect size was small (equivalent to a Cohen's $d=.33$). People were also more willing to consider the human an artist than the robot ($F(1,253)=99.789$, $p<.001$, $\eta_p^2=.283$), and here the effect size was very large (equivalent to $d=1.18$). Participants were significantly more willing to attribute mental states to the human than to the robot (*intention*, $p<.001$, $\eta_p^2=.086$; *desire*, $p<.001$, $\eta_p^2=.215$; *belief*, $p<.001$, $\eta_p^2=.178$).

---

[1] https://aspredicted.org/blind.php?x=qf6bu4. Experiments 1 and 2 were preregistered together and run in a single Qualtrics study preventing multiple participation. That way, people could not take part in both studies. Stimuli and data are available on OSF, see https://osf.io/huxq2/.

Aggregating across the two agent type conditions, participants were more inclined to judge the painting that resulted from intentional action to be art ($F(1,253)=6.582$, $p=.011$, $\eta_p^2=.025$), and they were more willing to judge the intentionally acting agent an artist ($F(1,253)=18.895$, $p<.001$, $\eta_p^2=.070$). No significant agent*behavior interaction was observed for the core DVs art or artistic agency. Expectedly, the interaction was significant for each of the three mental state DVs (see Appendix, Table A1). Figure 1 graphically presents the results in detail.

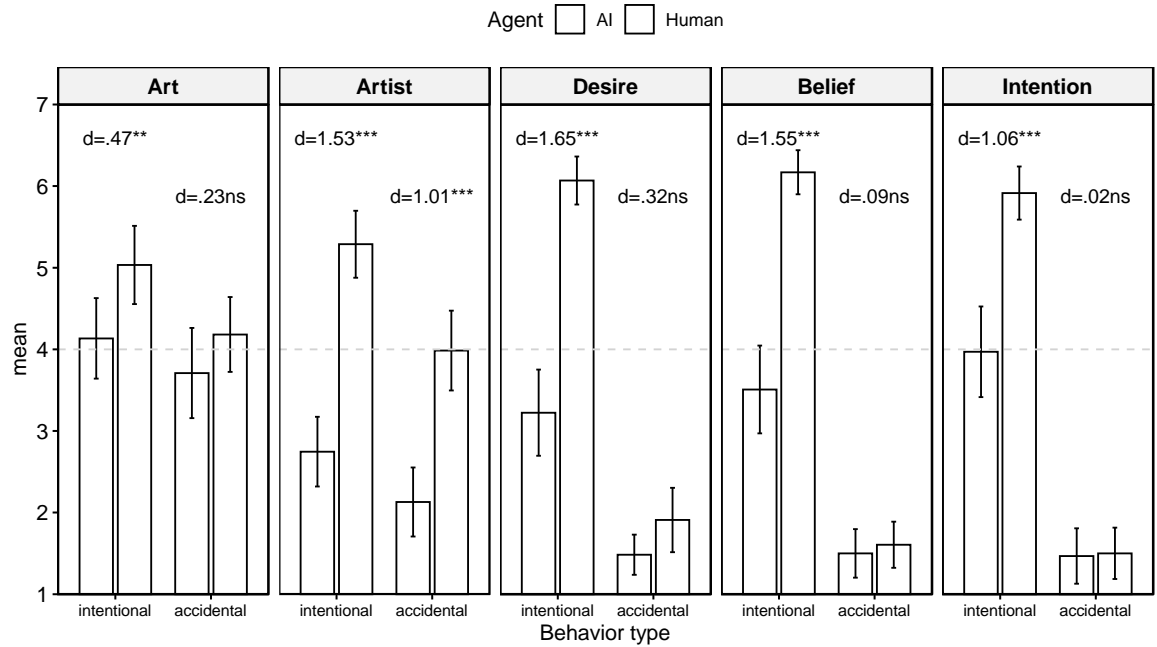

**Figure 1:** Mean ratings for art, artist, desire, belief, and intention across agent type (AI v. human agent) and behavior type (intentional v. accidental). Error bars denote 95%-confidence intervals. *$p<.05$; **$p<.01$; ***$p<.001$.

### 2.3.2   Explanatory DVs – Correlations.

The correlations between the core DVs – whether the painting was art and made by an artist – and the mental state DVs (desire, belief, intention) were all significant (all $ps<.001$, see Table 1). This suggests that the more pronounced the perceived desire, belief and intention to make a painting, the more pronounced the willingness to deem it art and its creator an artist. The correlation coefficients were considerably higher for the relations between mental states and artistic agency than for the relations between mental states and art judgments. Analyzing the data separately for each agent type produces similar-sized correlations for the AI and human agent subgroups (see Appendix, Table A3 and Table A4), though they tended to be somewhat stronger for the latter.

**Table 1: Bivariate correlations of DVs overall, two-tailed, \*sig.<.05, \*\*sig.<.01, \*\*\*sig.<.001.**

| r (Pearson) | Artist | Art | Desire | Belief | Intention |
|---|---|---|---|---|---|
| Artist | | .59*** | .53*** | .48*** | .42*** |

| r (Pearson) | Artist | Art | Desire | Belief | Intention |
|---|---|---|---|---|---|
| Art | .59*** | | .35*** | .35*** | .35*** |
| Desire | .53*** | .35*** | | .93*** | .81*** |
| Belief | .48*** | .35*** | .93*** | | .86*** |
| Intention | .42*** | .35*** | .81*** | .86*** | |

## 2.4 Discussion

Our experiment produced several findings: (i) Quite expectedly, people were more willing to deem an object art, and an agent an artist, if the latter acted intentionally rather than accidentally. This finding is broadly consistent with Levinson's [47, 48, 49, 50, 51] intention-dependent definition of art. Note, however, that the role of intentionality is perhaps less pronounced than one might assume. Aggregating across agents, the effect size of behavior type was small for art ($d$=.30) and medium-small for artistic agency ($d$=.41). (ii) Although we found that people are more willing to confer art-like status to the human's painting than to the robot's, the effect size of the difference was small (Cohen's $d$=.33, aggregated across behavior types). This suggests that people are quite willing to view robot art as art. By contrast, (iii) they are *much* less willing to consider robot agents artists than humans (here the effect size was very large, Cohen's $d$=1.18, aggregated across behavior types). This is likely due to the fact that, for artistic agency, perceived mental states play a more important role than for the status of art (see Table 1), and mental state ascriptions were significantly lower for the robot than for the human agent (aggregating across behavior types, belief: $d$=.68, desire: $d$=.58, intention: $d$=.34).

Given previous research concerning the folk's willingness to ascribe mental states to robot agents in moral contexts [85, 45, 81] and beyond [83, 23], we were somewhat surprised by the relatively low rates of their attribution in the robot conditions. In the intentional action condition, mean ascription of belief, desire and intention to the human agent were all significantly above the midpoint of the scale (one sample t-tests, all $ps$<.001, see Appendix Table A7), whereas they were not significantly above the midpoint for the robot. This, we reasoned, might be due to the fact that the vignette involved an *abstract* painting, rather than a representational painting: Due to the painting's abstract nature, people might have inferred that the robot lacked a genuine intention to make a painting and perhaps just splashed around with paint. Our hypothesis that artwork type may influence participants' judgments is supported by Chamberlain et al.'s [12] results, which show that people have a tendency to believe that artworks created by a computer are abstract. To explore this matter in more detail we ran a second experiment in which the created object is not an abstract painting, but a painting described as looking like a "relatively realistic representation of the local landscape".

## 3 Experiment 2 – Representational Art

### 3.1 Participants

We recruited 301 participants online via Amazon Mechanical Turk. The IP address location was restricted to the USA. In line with the preregistered criteria,[2] we excluded participants who failed an attention check or took less than eight seconds to answer the questions (not including reading the scenario), leaving a sample of 257 participants (female: 46 %; age M=43 years, SD=13 years, range: 22–88 years).

### 3.2 Methods and Materials

The methods and materials were identical to Experiment 1 except for features of the resulting painting. The painting was not described as "an abstract painting" but as looking like "a relatively realistic representation of the local landscape". There were again four conditions (2 agent types: human v. robot x 2 behavior types: intentional v. accidental), to which participants were randomly assigned. All vignettes are reported in detail in the Appendix.

---

[2] https://aspredicted.org/blind.php?x=qf6bu4. Stimuli and data are available on OSF, see https://osf.io/huxq2/.

## 3.3 Results

### 3.3.1 Main Results.

As in Experiment 1, a series of mixed-design ANOVAs determined that, aggregating across the two behavior type conditions, participants were more willing to consider the human an artist than the robot ($F(1,103)=107.353$, $p<.001$, $\eta_p^2=.297$, equivalent to a Cohen's d of 1.19, a large effect). Importantly, however, there was no significant difference in the willingness of the participants to confer the status of art to the human's painting than to the robot's ($F(1,256)=.534$, $p=.466$, $\eta_p^2=.002$, equivalent to a Cohen's d=.09). Participants were significantly more willing to attribute mental states to the human than to the robot (intention, $p<.001$, $\eta_p^2=.044$; desire, $p<.001$, $\eta_p^2=.109$; belief, $p<.001$, $\eta_p^2=.099$). However, the effect sizes were only about half as pronounced as in the abstract painting experiment.

Aggregating across the two agent type conditions, participants were more inclined to judge the painting that resulted from intentional action to be art ($F(1,256)=13.081$, $p<.001$, $\eta_p^2=.050$), and they were more willing to judge the intentionally acting agent an artist ($F(1,256)=46.940$, $p<.001$, $\eta_p^2=.159$). No significant agent*behavior interaction was observed for the core DVs art or artistic agency. Expectedly, the interaction was significant for each of the three mental state DVs (see Appendix, Table A2). Figure 2 graphically presents the results in detail.

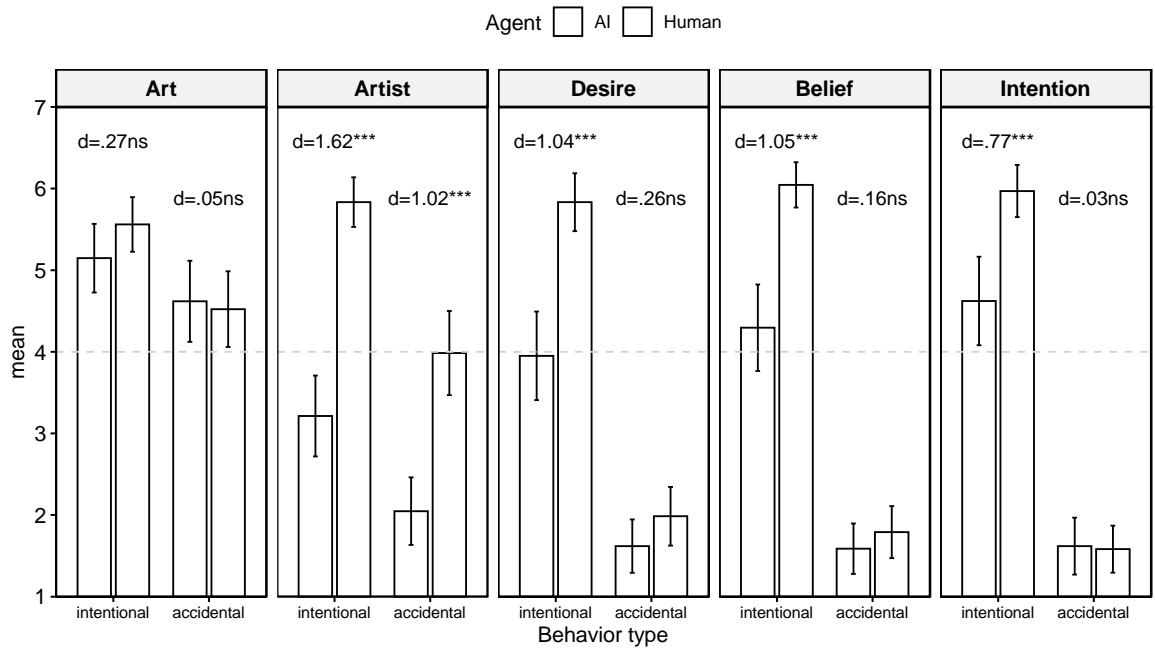

**Figure 2:** Mean ratings for art, artist, desire, belief, and intention across agent type (AI v. human agent) and behavior type (intentional v. accidental). Error bars denote 95%-confident intervals. *$p<.05$; **$p<.01$; ***$p<.001$.

### 3.3.2 Explanatory DVs – Correlations.

Replicating the findings from Experiment 1, the correlations between the core DVs – whether the painting was art and made by an artist – and the mental state DVs (desire, belief, intention) were all significant (all $ps<.01$, see Table 2). This again suggests that the more pronounced the perceived desire, belief and intention to make a painting, the more pronounced the willingness to deem it art and its creator an artist. The correlation coefficients were considerably higher for the relations between mental states and artistic agency than for the relations between mental states and art judgments. Analyzing the data separately for each agent type produces similar-sized correlations for the AI and human agent subgroups (see Appendix, Table A5 and Table A6), though they tended to be somewhat stronger for the latter.

**Table 2: Bivariate correlations of DVs overall, two-tailed, *sig.<.05, **sig.<.01, ***sig.<.001.**

| r (Pearson) | Artist | Art | Desire | Belief | Intention |
|---|---|---|---|---|---|
| Artist |  | .49*** | .55*** | .51*** | .49*** |
| Art | .49*** |  | .36*** | .35*** | .40*** |
| Desire | .55*** | .36*** |  | .90*** | .81*** |
| Belief | .51*** | .35*** | .90*** |  | .90*** |
| Intention | .49*** | .40*** | .81*** | .90*** |  |

### 3.3.3 Joint Analyses.

We conducted joint analyses for the entire sample of the two experiments (n=511, female: 48 %; mean age: 43 years, SD = 14 years, range: 20–88 years). This allowed us to make *object type* (abstract v. representational painting) a third factor beyond *agent type* (human vs. AI) and *behavior type* (intentional vs. accidental). A series of mixed-design ANOVAs showed that, aggregating across behavior type and agent type, participants were significantly more willing to confer the status of art to the representational painting than to the abstract painting ($F(1,510)=17.910$, $p<.001$, $\eta_p^2=.034$). No other significant main effects or interactions were observed for object type (see Table 3), except for the agent*behavior interaction for artistic agency, where the effect size, however, was very small ($\eta_p^2=.009$).

**Table 3: Threeway-ANOVAs, two-tailed, *sig.<.05, **sig.<.01, ***sig.<.001.**

| $\eta_p^2$ (N=511) | Art | Artist | Intention | Desire | Belief |
|---|---|---|---|---|---|
| Agent | .013* | .291** | .065** | .161** | .139** |
| Behavior | .036** | .111** | .569** | .480** | .571** |
| Object | .034** | .044 | .006 | .003 | .006 |
| Agent*Behavior | .004 | .009* | .065** | .089** | .109** |
| Agent*Object | .005 | .000 | .003 | .007 | .005 |
| Behavior*Object | .000 | .006 | .001 | .001 | .001 |
| Agent*Behavior*Object | .000 | .000 | .002 | .005 | .007 |
| adjusted $R^2$ | .074** | .347** | .588** | .541** | .612** |

## 3.4 Discussion

Experiment 2 replicated several core findings of Experiment 1: (i) Consistent with Levinson's account of art [47, 48, 49, 50, 51], people were more willing to deem an object art, and an agent an artist, if the latter acted purposefully rather than accidentally. In contrast to Experiment 1, the impact of intentionality on both cored DVs was somewhat more pronounced. (ii) Once again, people were much less willing to consider robot agents artists than humans (aggregating across behavior types, the effect size was again very large, $d=1.19$). As in Experiment 1, perceived mental states correlated more strongly with artistic agency than with an object's being considered art.

Importantly, and in line with our hypothesis, the nature of the painting (abstract v. representational) did have some impact: Whereas people were less willing to view the robot painting as art than the human painting in Experiment 1, no significant difference could be found across agent types in Experiment 2. Differently put, people were *just as willing to consider the robot's painting as art as the human's*. As hypothesized, the ascriptions of mental states to the robot were higher in the representational painting conditions than in the abstract painting conditions. However, the same held for the human agent. While the difference in perceived belief, desire and intention remained significant across agent types in Experiment 2, the effect sizes were considerably smaller than in Experiment 1. This is presumably what explains why the paintings of human and robot were deemed art to more similar extents in Experiment 2 than in Experiment 1.

# 4 GENERAL DISCUSSION

In two experiments concerning an object's status as art and its creator's status as an artist, we manipulated three factors: (i) behavior type (intentional v. accidental), (ii) agent type (AI-driven robot v. human) and (iii) object type (abstract v. representational painting). The first factor tells us something about the folk concepts of art and artistic agency broadly conceived (i.e., not limited to robot-art). Artistic agency seems to be quite strongly tied to an intention of making a work of art. Whether the work constitutes art is also significantly impacted by whether it was made intentionally. However, the results for Experiment 2 show at least moderate agreement with the claim that accidentally produced representational paintings *can* constitute art. This suggests that intentionality is a relevant criterion for an object's status as art, though (and *pace* Levinson [47], quoted above) it also suggests that, on the folk view, intentions do not constitute a *necessary* condition for art. Given this finding, the folk concept of art thus allows for objects to be considered art which have been created by agents that might be incapable of having full-fledged artistic intentions.

The manipulation of the second factor – agent type – confirms this. Although people were not particularly willing to ascribe mental states to the robot agent (at least in comparison to the results reported in [85, 53, 81, 82]), they were nonetheless rather willing to consider the robot paintings art (when not created by accident). Averaging across behavior types, we found no significant difference in the attribution of the status of art across agent types (robot v. human) for the representational painting (Experiment 2), and for the abstract painting (Experiment 1) the effect size was small. In other words, the folk are by and large as willing to consider robot creations as art as human creations (if all else is held fixed). Interestingly, however, this perceived similarity does not extend to creative agency: Robots whose paintings are deemed art are not considered artists, whereas humans are – and here the effect-size of the difference was very large in both experiments.

The third factor – abstract v. representational painting – had a significant main effect on the ascription of the status of art. It did not have a main effect on any of the other dependent variables, and all interactions with object type were nonsignificant for all five DVs.

Our experiments explored three core questions: Whether people deem robot creations art, consider AI-driven robots artists and are willing to attribute mental states to them. We will discuss them in reverse order. As concerns mental states, we made two astonishing findings. First, in contrast to moral contexts and a general tendency to anthropomorphize robots [77], people are relatively unwilling to attribute desire, belief and intentions to artificial agents in aesthetic contexts. These unexpected differences of mental state ascriptions across domains invite further research. A second finding of considerable importance regards the role of intentions in artistic creation in general (i.e. independently of AI). A plethora of studies in psychology of art have emphasized the importance of the creator's intentions in the artistic process [40, 65, 70, 5, 32, 33, 71]. Our study confirms that the intentions of the artist constitute a core factor in judgments about what is art and what is not. However, we found, accidental creations produced without artistic intentions are sometimes also deemed art.

This result, i.e. that for an object to be considered art it does not *necessarily* have to be created intentionally, is also important from the perspective of philosophy of art. To date, there exists very little research on the folk concept of art (Kamber conducted two studies to test the intuitiveness of various definitions of art [41, 42]; another study by Liao et al. has shown that the folk concept of art is a dual-character concept [52]). Future research in experimental aesthetics might explore the folk concept of art and its relation to intentional creation more thoroughly. Potential factors of interest to manipulate in future studies include, for instance, (i) different types of artworks – music, poems, literature etc. (experimental philosophy of music [6, 60, 61, 62], for instance, would be enriched by research on robot composers), (ii) distinct types of qualities of the artwork that figure prominently in certain definitions of art such as, for instance, beauty [7] or institutional recognition [25, 21].

As regards the other two core questions, we found that, on the folk view, robots *can* make art though they are *not* considered artists. This interesting result raises a series of discussion points. One regards the anthropomorphism of robot agents. We know from previous research that people tend to value robot-created art less than human art (both visual art [43, 12, 72] and music [63], though see [38]), and that this tendency decreases when the robot appears to be more anthropomorphic [12]. These results are consistent with broader research about robot anthropomorphism, which shows more favourable attitudes towards robots with human-like traits (be it empathy towards them [73], felt proximity [31], social attitudes toward robots [83], or the assessment of their moral qualities [56] or "personality" [11]). Future research could explore whether human-like features of the robot impact our DVs (art, artistic agency, mental states) in the context of art creation. This kind of research would also be important for robot designers and developers who may want to encourage (or discourage) people's tendency to anthropomorphise robots in the context of art.

Our rather foundational inquiry into the folk concepts of art and artistic agency could be brought to bear on more pragmatic – yet by no means less interesting – questions, such as who gets the credit for, or holds the copyright of, art in the production of which AI was partially involved (see e.g. [30]). Our current results, which show that people do not see robots as artists, lead us to predict that people will not be very willing to recognise AI as potential artwork copyright holders. Somaya and Varshney, however, note that AI design, physical embodiment, and anthropomorphism in particular might influence these judgments [78]. In future studies, it might also be interesting to contrast robot agents not only with singular human agents, but group agents such as art collectives, or not yet fully formed human agents such as small children, or human-robot teams.

Finally, there are some limitations to our studies. First, following related work in experimental philosophy [41, 42, 52, 18, 19, 20], we opted for a single-item approach to measure the "art" and "artist" perceptions. Studies of this sort might benefit from the use of more complex multi-item scales (see e.g. [38]). Second, as philosophers, we are principally interested in the underexplored *folk concepts* of art and artistic agency. Naturally, contrasting the latter with results from expert samples (e.g. artists, art historians, or philosophers of art) might constitute a worthwhile endeavor. Third, we have only collected and analyzed quantitative data. A mixed-methods approach in which participants are encouraged to justify their assessments might deliver further valuable insights.

## ACKNOWLEDGMENTS

We would like to thank Jerrold Levinson for his helpful comments, Marc-André Zehnder and the Digital Society Initiative (University of Zurich) for assistance and support, and the Swiss National Science Foundation for funding (Grant no: PZ00P1_179912, PI: Kneer).

# APPENDIX
## 1 Vignettes
### 1.1 Experiment 1.

**Condition 1 (human/intentional/abstract):** Imagine a person in an art studio who decides to create a piece of art. She takes an empty canvas and starts splashing paint onto it. In the end, the object looks like an abstract painting.

**Condition 2 (human/accidental/abstract):** Imagine a person in an art studio who is currently tidying up. She accidentally brushes against some jars of paint that spill onto an empty canvas. In the end, the object looks like an abstract painting.

**Condition 3 (AI/intentional/abstract):** Imagine a robot equipped with artificial intelligence that can make decisions autonomously. The robot is in an art studio and decides to create a piece of art. It takes an empty canvas and starts splashing paint onto it. In the end, the object looks like an abstract painting.

**Condition 4 (AI/accidental/abstract):** Imagine a robot equipped with artificial intelligence that can make decisions autonomously. The robot is in an art studio and currently tidying up. It accidentally brushes against some jars of paint that spill onto an empty canvas. In the end, the object looks like an abstract painting.

### 1.2 Experiment 2.

**Condition 1 (human/intentional/representational):** Imagine a person in an art studio who decides to create a piece of art. She takes an empty canvas and starts splashing paint onto it. In the end, the object looks like a relatively realistic representation of the local landscape.

**Condition 2 (human/accidental/representational):** Imagine a person in an art studio who is currently tidying up. She accidentally brushes against some jars of paint that spill onto an empty canvas. In the end, the object looks like a relatively realistic representation of the local landscape.

**Condition 3 (AI/intentional/representational):** Imagine a robot equipped with artificial intelligence that can make decisions autonomously. The robot is in an art studio and decides to create a piece of art. It takes an empty canvas and starts splashing paint onto it. In the end, the object looks like a relatively realistic representation of the local landscape.

**Condition 4 (AI/accidental/representational):** Imagine a robot equipped with artificial intelligence that can make decisions autonomously. The robot is in an art studio and currently tidying up. It accidentally brushes against some jars of paint that spill onto an empty canvas. In the end, the object looks like a relatively realistic representation of the local landscape.

## 2. ANOVAs
### 2.1 ANOVA Table for Experiment 1.

**Table A1:** Twoway-ANOVAs for the abstract painting, two-tailed, *sig.<.05, **sig.<.01, ***sig.<.001.

| $\eta_p^2$ (N=254) | Art | Artist | Intention | Desire | Belief |
|---|---|---|---|---|---|
| Agent | .029** | .283*** | .086*** | .215*** | .178*** |
| Behavior | .025* | .070*** | .537*** | .472*** | .550*** |
| Agent*Behavior | .003 | .010 | .082*** | .132*** | .158** |
| adjusted $R^2$ | .042** | .301*** | .562*** | .555*** | .608*** |

*2.2 ANOVA Table for Experiment 2.*

**Table A2:** Twoway-ANOVAs for the representational painting, two-tailed, *sig.<.05, **sig.<.01, ***sig.<.001.

| $\eta_p^2$ (N=257) | Art | Artist | Intention | Desire | Belief |
|---|---|---|---|---|---|
| Agent | .002 | .297*** | .044*** | .109*** | .099*** |
| Behavior | .050*** | .159*** | .601*** | .489*** | .592*** |
| Agent*Behavior | .005 | .009 | .050*** | .054*** | .066*** |
| adjusted $R^2$ | .046** | .377*** | .612*** | .527*** | .616*** |

## 3. Correlations

*3.1 Further Correlations for Experiment 1.*

**Table A3:** Bivariate correlations of DVs in the AI group, two-tailed, *sig.<.05, **sig.<.01, ***sig.<.001.

| r (Pearson) | Artist | Art | Desire | Belief | Intention |
|---|---|---|---|---|---|
| Artist |  | .62*** | .43*** | .43*** | .37*** |
| Art | .62*** |  | .37*** | .34*** | .36*** |
| Desire | .43*** | .37*** |  | .93*** | .75*** |
| Belief | .43*** | .34*** | .93*** |  | .77*** |
| Intention | .37*** | .36*** | .75*** | .77*** |  |

**Table A4:** Bivariate correlations of DVs in the Human group, two-tailed, *sig.<.05, **sig.<.01, ***sig.<.001.

| r (Pearson) | Artist | Art | Desire | Belief | Intention |
|---|---|---|---|---|---|
| Artist |  | .59*** | .46*** | .42*** | .41*** |
| Art | .59*** |  | .29** | .32*** | .32*** |
| Desire | .46*** | .29** |  | .92*** | .86*** |
| Belief | .42*** | .32*** | .92*** |  | .93*** |
| Intention | .41*** | .32*** | .86*** | .93*** |  |

*3.2 Further Correlations for Experiment 2.*

**Table A5:** Bivariate correlations of DVs in the AI group, two-tailed, *sig.<.05, **sig.<.01, ***sig.<.001.

| r (Pearson) | Artist | Art | Desire | Belief | Intention |
|---|---|---|---|---|---|
| Artist |  | .45*** | .43*** | .42*** | .39*** |
| Art | .45*** |  | .23** | .28** | .34*** |
| Desire | .43*** | .23** |  | .84*** | .71*** |
| Belief | .42*** | .28** | .84*** |  | .85*** |
| Intention | .39*** | .34*** | .72*** | .85*** |  |

**Table A6:** Bivariate correlations of DVs in the Human group, two-tailed, *sig.<.05, **sig.<.01, ***sig.<.001.

| r (Pearson) | Artist | Art | Desire | Belief | Intention |
|---|---|---|---|---|---|
| Artist |  | .63*** | .58*** | .53*** | .57*** |
| Art | .63*** |  | .46*** | .42*** | .45*** |
| Desire | .58*** | .46*** |  | .93*** | .89*** |
| Belief | .53*** | .42*** | .93*** |  | .93*** |
| Intention | .57*** | .45*** | .89*** | .93*** |  |

## 4. One-sample t-tests

**Table A7:** One-sample t-tests with H0: mean=4.

| Object | Agent type | Behavior type | DV | n | t(n) | p | Bonferroni adjusted p |
|---|---|---|---|---|---|---|---|
| abstract | AI | accidental | Art | 62 | -1.05 | .300 | >.999 |
| abstract | AI | accidental | Artist | 62 | -8.85 | <.001 | <.001 |
| abstract | AI | accidental | Desire | 62 | -20.41 | <.001 | <.001 |
| abstract | AI | accidental | Belief | 62 | -16.83 | <.001 | <.001 |
| abstract | AI | accidental | Intention | 62 | -14.89 | <.001 | <.001 |
| abstract | AI | intentional | Art | 67 | .54 | .590 | >.999 |
| abstract | AI | intentional | Artist | 67 | -5.86 | <.001 | <.001 |
| abstract | AI | intentional | Desire | 67 | -2.93 | .010 | .184 |
| abstract | AI | intentional | Belief | 67 | -1.83 | .070 | >.999 |
| abstract | AI | intentional | Intention | 67 | -.11 | .920 | >.999 |
| abstract | Human | accidental | Art | 66 | .79 | .430 | >.999 |

| Object | Agent type | Behavior type | DV | n | t(n) | p | Bonferroni adjusted p |
|---|---|---|---|---|---|---|---|
| abstract | Human | accidental | Artist | 66 | -.06 | .950 | >.999 |
| abstract | Human | accidental | Desire | 66 | -10.58 | <.001 | <.001 |
| abstract | Human | accidental | Belief | 66 | -16.93 | <.001 | <.001 |
| abstract | Human | accidental | Intention | 66 | -15.87 | <.001 | <.001 |
| abstract | Human | intentional | Art | 59 | 4.32 | <.001 | .002 |
| abstract | Human | intentional | Artist | 59 | 6.28 | <.001 | <.001 |
| abstract | Human | intentional | Desire | 59 | 14.09 | <.001 | <.001 |
| abstract | Human | intentional | Belief | 59 | 16.08 | <.001 | <.001 |
| abstract | Human | intentional | Intention | 59 | 11.77 | <.001 | <.001 |
| representational | AI | accidental | Art | 63 | 2.49 | .020 | .614 |
| representational | AI | accidental | Artist | 63 | -9.45 | <.001 | <.001 |
| representational | AI | accidental | Desire | 63 | -14.53 | <.001 | <.001 |
| representational | AI | accidental | Belief | 63 | -15.61 | <.001 | <.001 |
| representational | AI | accidental | Intention | 63 | -13.65 | <.001 | <.001 |
| representational | AI | intentional | Art | 61 | 5.46 | <.001 | <.001 |
| representational | AI | intentional | Artist | 61 | -3.18 | <.001 | .093 |
| representational | AI | intentional | Desire | 61 | -.18 | .860 | >.999 |
| representational | AI | intentional | Belief | 61 | 1.11 | .270 | >.999 |
| representational | AI | intentional | Intention | 61 | 2.3 | .030 | .996 |
| representational | Human | accidental | Art | 67 | 2.25 | .030 | >.999 |
| representational | Human | accidental | Artist | 67 | -.06 | .950 | >.999 |
| representational | Human | accidental | Desire | 67 | -11.21 | <.001 | <.001 |
| representational | Human | accidental | Belief | 67 | -13.81 | <.001 | <.001 |
| representational | Human | accidental | Intention | 67 | -16.74 | <.001 | <.001 |
| representational | Human | intentional | Art | 66 | 9.32 | <.001 | <.001 |
| representational | Human | intentional | Artist | 66 | 12.06 | <.001 | <.001 |
| representational | Human | intentional | Desire | 66 | 10.33 | <.001 | <.001 |
| representational | Human | intentional | Belief | 66 | 14.72 | <.001 | <.001 |
| representational | Human | intentional | Intention | 66 | 12.3 | <.001 | <.001 |

18